\documentclass[12pt,preprint]{aastex}

\shorttitle{Massive Protoplanetary Disks in Orion}
\shortauthors{Mann \& Williams}

\received{2009}
\begin{document}

\title{Massive Protoplanetary Disks in Orion Beyond the Trapezium Cluster}
\author{Rita K. Mann, \and Jonathan P. Williams}

\affil{Institute for Astronomy, University of Hawaii, 2680 Woodlawn Drive, Honolulu, HI 96822}
\email{rmann@ifa.hawaii.edu, jpw@ifa.hawaii.edu}

\begin{abstract}
We present Submillimeter Array\footnote{The Submillimeter Array is a joint
project between the Submillimeter Astrophysical Observatory and the Academica
Sinica Institute of Astronomy and Astrophysics and is funded by the Smithsonian
Institution and the Academica Sinica.} observations of the $880\,\mu$m
continuum emission from three circumstellar disks around young stars in Orion that
lie several arcminutes ($\gtrsim$\,1-pc) north of the Trapezium cluster. 
Two of the three disks are in the binary system 253-1536.
Silhouette disks 216-0939 and 253-1536a are found to be more massive than any
previously observed Orion disks, with dust masses derived from their submillimeter 
emission of $0.045\,M_\odot$ and $0.066\,M_\odot$, respectively.
The existence of these massive disks reveals the disk mass distribution
in Orion does extend to high masses, and that the truncation observed in the
central Trapezium cluster is a result of photoevaporation due to the proximity
of O-stars.  253-1536b has a disk mass of $0.018\,M_\odot$, making the
253-1536 system the first optical binary in which
each protoplanetary disk is massive enough to potentially form Solar systems.
\end{abstract}

\keywords{circumstellar matter --- planetary systems: protoplanetary disks 
--- solar system: formation --- stars: pre-main sequence}

\section{Introduction}\label{sec: intro}
The formation of planetary systems is intimately connected to the
properties of the circumstellar disks in which they are born.
Most studies to date have focused on young disks in Taurus-Auriga and 
$\rho$ Ophiuchus for their proximity.  However, most stars do not form 
in relative isolation as in Taurus and $\rho$ Ophiuchus, but rather in 
dense, massive star clusters that produce ionizing radiation \citep{lada}, which
threatens the persistence of nearby circumstellar disks \citep{johnstone98}. 
There is even clear evidence
based on the presence of short-lived radionucleides, particularly $^{60}$Fe,
in meteorites \citep{tachibana}, which indicate our own Solar System
originated in a massive star forming environment \citep{krot,gaidos}.

Orion is the nearest young massive star forming region, and its favourable
geometry due to a ``blister HII'' region aligned toward the Sun has allowed
over 200 young protoplanetary disks (``proplyds'') to be discovered in projection
against the bright background nebula by the 
Hubble Space Telescope (HST) \citep{odell94,odell96,bally98a,bally00,ricci}.
The potential for the proplyds to form planetary systems depends on how
much mass remains in the disks, in spite of the hostile environment.
Several millimeter interferometric surveys have been undertaken to determine
the masses of these disks, but few proplyds were detected in excess of 
the ionized gas emission, which swamps the dust emission at centimeter to 
millimeter wavelengths \citep{mundy,bally98b,eisner06,eisner08}.
With its higher frequency observations, the Submillimeter Array (SMA) 
is more sensitive to dust-disk emission from the Orion proplyds than any 
existing interferometer \citep{williams}.

The initial results of our SMA survey of disk masses in the Orion Trapezium
cluster revealed this region is missing the most massive disks found in 
Taurus and $\rho$\,Ophiuchus, most likely a result of photoevaporation
by the most massive star of the cluster, $\theta^1$\,Ori\,C 
(\citet{mann}, hereafter Paper I).  Photoevaporation
rapidly erodes large and massive disks near O-stars, but at
larger distances, the flux of ionizing radiation drops significantly and
no longer influences the disks.
Using our observations, we searched for a disk mass dependence on distance
from $\theta^1$\,Ori\,C, but did not find one.  In order to explore
the extent of the disk mass truncation in Orion, we imaged two HST 
identified proplyds at larger distances from the Trapezium cluster.

In this Letter, we present our submillimeter observations of two HST-identified disks,
216-0939 and 253-1536\footnote{Proplyd designation based on the nomenclature of \cite{odell94}},
located in the outskirts of the Orion Nebula (see Figure 1).
These disks are seen in silhouette and were discovered by
\citet{smith}.  Proplyd 216-0939 is a 2.6$^{''}$ flared disk surrounding a
$0.69\,M_\odot$ star (Hillenbrand 1997), while proplyd 253-1536 is a suspected
binary system \citep{nielbock,smith,kohler,reipurth}, with a disk
surrounding the apparently fainter companion.  Both disks lie
many arcminutes ($\gtrsim$\,1-pc) North of the Trapezium cluster (see Figure 1),
well away from the ionizing radiation of $\theta^1$ Ori C.
The SMA observations are described in Section \S2, followed by our determination 
of disk masses in Section \S3.  We discuss the implications of our observations 
for planet formation in Orion in Section \S4.

\section{Observations}
Submillimeter interferometric observations at $880\,\mu$m were conducted 
with the Submillimeter Array (SMA) in its compact configuration on 2008 
Dec 23 and 2008 Dec 24 on Mauna Kea towards disks 216-0939 and 253-1536.  
In this arrangement, the SMA's eight 6-m antennas provide a maximum baseline 
of 70-m.  Additional observations of 253-1536 were obtained on 2009 March 26 
and 27 using the very extended (VEX) configuration, which provides baselines 
up to 508-m.  Double sideband receivers were tuned to an intermediate frequency 
(IF) of 340.175 GHz.  
Each sideband provides 2\,GHz of bandwidth, separated by $\pm$\,5\,GHz from the IF.  
We simultaneously observed the CO(3--2) transition, which was strongly
detected but maps show any line emission from the disk is substantially contaminated
by confusion with the more extended molecular cloud background.
Weather conditions for the observations were good, with $<$\,3-mm precipitable
water vapor, resulting in system temperatures ranging from 100-400\,K.

The raw visibilities for each night were calibrated and edited using the 
MIR software package.  The phase and amplitudes were monitored with 5 minute 
integrations of two quasars (J0423--013 and J0530+135), which were interleaved 
with 20 minute on-source integrations.  Passband calibration was conducted with 
the bright quasar 3c273 and Uranus was used to set the absolute flux scale, 
which is accurate to $\sim$ 10\%. The calibrated visibilities were weighted 
by system temperature and inverted, then deconvolved to generate continuum 
maps using the MIRIAD software package.

CO(3--2) was edited out of the $880 \,\mu$m observations to generate a 
line-free continuum, which was used to produce the final maps shown in Figure 2.
Images were created for uv-spacings greater than 27\,k$\lambda$, or physical 
baselines longer than 23-m, to filter out uniform extended emission greater 
than 7.5$\arcsec$ in size, larger than resolved HST disk sizes.

The map of 216-0939 was made with only compact array observations, so
longer baseline data was weighted higher (super-uniform weighting) in order to 
maximize the resolution to get a circular $1.4\arcsec\times\,1.4\arcsec$
beam.  The map of 253-1536 combined the compact and VEX SMA observations.
The addition of VEX observations to the compact configuration
data significantly improved the beam size from $1.5\arcsec\,\times\,1.4\arcsec$
to $0.3\arcsec \times 0.2\arcsec$. 
The improvement in resolution allowed us to resolve and distinguish disk emission from
each source in the binary, and we label the primary and secondary disks
253-1536a and 253-1536b, respectively.

\section{Results}
Figure 2 shows the $880\,\mu$m continuum images of proplyds 
216-0939 and 253-1536 alongside their discovery HST H$\alpha$ images
taken from \citet{smith}.  All three disks were very strongly detected,
and their integrated flux densities are listed in Table 1.
 
As the silhouette disks show no signs of being photoevaporated, the free-free
emission from  
ionized gas emission is not expected to be significant.  The disks lie in regions 
where the molecular cloud emission is lower and more uniform than for 
the central Trapezium cluster surveyed in Paper I.  Bolometer maps made
with the SCUBA camera on the James Clerk Maxwell Telescope by
\citet{johnstone99} were used to correct background emission in a similar
way to Paper I.  The background was found to contribute
$<\arrowvert$1\,mJy$\arrowvert$ to the disk fluxes, 
much less than calibration errors.
Therefore, since both the ionized gas and background molecular cloud emission are 
negligible, the $880\,\mu$m fluxes arise entirely from the dust-disks
and unlike Paper I, we do not make any corrections.

The disk masses were derived from the dust fluxes using the
standard relationship from \citep{beckwith}:

\begin{equation} \label{equ: masses}
M_{\rm disk} = \frac{F_{\rm dust}d^2}{\kappa_{\nu}B_{\nu}(T)},
\end{equation}

where $d=400$\,pc is the distance to Orion \citep{sandstrom,menten},
$\kappa_{\nu}=0.1(\nu/1000\,{\rm GHz})=0.034\,{\rm cm}^2\,{\rm g}^{-1}$
is the dust grain opacity with an implicit gas-to-dust mass ratio of 100:1,
and $B_{\nu}(T)$ is the Planck function.  A dust temperature of $T=20$\,K
was used, which is the average for disks in Taurus-Auriga and $\rho$\,Ophiuchus 
\citep{andrews05,andrews07}.  We used the same temperature and opacity as these
disk surveys and also of Orion \citep{mann} for consistency.

The resulting disk masses are $0.045\,M_\odot$ for proplyd 216-0939,
$0.066\,M_\odot$ for 253-1536a and $0.018\,M_\odot$ for 253-1536b (see Table 1).  
Both 216-0939 and 253-1536a, are more massive than the upper limit 
of $0.034\,M_\odot$ found for the Orion Trapezium cluster proplyds in Paper I.

\begin{deluxetable}{lcccccr}
\tablecolumns{7}
\tablewidth{0pc}
\tabletypesize{\scriptsize}
\tablecaption{Disk Fluxes and Masses\label{}}
\tablehead{
\colhead{Proplyd} & \colhead{$F_{\rm obs}$} & \colhead{rms} &
\colhead{$M_{\rm disk}$} & \colhead{Disk Radius} & \colhead{SpT}\\ 

\colhead{Name} & \colhead{(mJy)} & \colhead{(mJy)} &
\colhead{($10^{-2}\,M_\odot$)}  & \colhead{(AU)} & \colhead{} \\

\colhead{(a)} & \colhead{(b)} & \colhead{(c)} & \colhead{(d)} 
              & \colhead{(e)} & \colhead{(f)} 
}
\startdata
216-0939 &  91.9  & 1.3 &  4.50 $\pm$ 0.06 & 291     & K5   \\
253-1536a & 134.2 & 1.0 &  6.62 $\pm$ 0.13 & 282     & unknown    \\
253-1536b & 37.5  & 1.0 &  1.84 $\pm$ 0.04 & $<$\,60 & M2.5 \\
\enddata
\tablecomments{
(a) Proplyd designation based on the nomenclature of \cite{odell94}.
(b) Integrated continuum flux density from the disk.
(c) 1$\sigma$ statistical error
(d) Disk mass (error does not include uncertainties in the flux scale of $\sim$ 10\%).
(e) Size of submillimeter emission determined using uvfit in MIRIAD.
(f) Spectral type of embedded star from \cite{hillenbrand}.
}
\end{deluxetable}

Both disks 216-0939 and 253-1536a are resolved. An elliptical gaussian fit 
to the visibilities gives a size of $1.5\arcsec \times 0.3\arcsec$ 
at a position angle of $174\arcdeg$ (measured east of north) for disk 216-0939, which corresponds
to a disk radius of $\sim 300$\,AU and inclination $\sim$\,80$\arcdeg$.
The fit to disk 253-1536a gives a size of $1.4\arcsec 
\times 0.6\arcsec$ at a position angle of 73$\arcdeg$, 
corresponding to a radius of 280AU and
inclination of $\sim$\,65$\arcdeg$.
The secondary disk, 253-1536b, is unresolved, implying a radius
$<$\,60\,AU.  

\section{Discussion}
\subsection{Massive Disks and Planet Formation Capability}
We found in Paper I that the disk mass distribution in the Trapezium
cluster is truncated, with a lack of massive disks (M$>0.034\,M_\odot$)
relative to Taurus and $\rho$\,Ophiuchus.  
As discussed in Paper I, the relatively low disk luminosities in the
center of the Trapezium cluster cannot be due to heating from the O
stars because that would operate in the opposite way to produce higher
submillimeter fluxes for a given dust mass.
The dust opacity in disks is unknown and a lower value in
the Trapezium cluster disks, due to substantial grain growth,
might explain the discrepancy\citep{pollack,ossenkopf}.
However, given that the cluster disks tend to have smaller sizes
than the massive Taurus disks and the objects in this
Letter, a more reasonable interpretation for the truncation of
the Trapezium cluster disk mass distribution is the loss
of the outer edges of the largest disks due to photoevaporation.
Our finding here, of more massive disks beyond the Trapezium
cluster, reinforces this interpretation. Proplyds 216-0939 and 253-1536
show no signs of photoevaporation and are similar in size and mass to
the largest disks seen in Taurus.

Stellar encounters are much more common in the Trapezium cluster
than in the more sparsely populated lower mass star forming regions.
Although \cite{scally} found that they are an insignificant
factor in disk destruction at the 1-2\,Myr age of the cluster,
\cite{olczak} reasoned otherwise based on more detailed
modeling of the disk-disk interaction and considering encounters with
massive stars. Our disk mass measurements provide an important
constraint on these theories. 
The most massive disk we have measured here, $0.066\,M_\odot$, is almost  
twice that of the most massive disk we observed in the Trapezium  
cluster, $0.034\,M_\odot$. If this discrepancy were only due to stellar  
encounters, it would require not only a close encounter but also a  
high stellar mass ratio: $<\,1000$\,AU for $M_2/M_1=90$ \citep{olczak}.
Yet, these same conditions imply a very high photoevaporative loss rate
due to an O star.
The low disk masses within the Trapezium cluster together with
the high disk masses at larger distances firmly establish photoevaporation
as the dominant disk erosive agent.

Counting 216-0939 and the two resolved disks in the binary 253-1536,
we have measured the masses of three disks. Unless they have very flat
surface density profiles, all satisfy the mass and
size requirement to form a planetary system on the scale of
our own; M$>0.01\,M_\odot$ within 60 AU (Paper I).
\cite{clarke} stated that the $880\,\mu$m emission may become
optically thick in such compact disks.  But for the \cite{beckwith}
dust opacity we have used here, $\tau=\kappa\Sigma\approx 0.1$,
where the average surface density, $\Sigma=4$ g\,cm$^{-2}$,
for a face-on disk with M=$0.01\,M_\odot$ and R=60AU.
The innermost regions may be optically thicker
if the surface density strongly increases and $\tau$ may approach
unity for edge-on disks such as 216-0939 but we did not see a
correlation between flux and disk orientation in Paper I and believe
that our observations are a good measure of the amount of small
grains in the disks.
There may well be undetected centimeter-sized and larger particles
in the proplyds (e.g., \cite{wilner}). These massive silhouette
disks at large distances from $\theta^1$\,Ori\,C have no detectable free-free
emission at the mJy level
and are good candidates for longer wavelength observations
to study the larger grain population.
It is important to keep in mind, of course, that the main uncertainty
in any mass estimate based on the dust continuum is the gas-to-dust ratio.

\subsection{The Binary Proplyd 253-1536}
The HST image of proplyd 253-1536a by \citet{smith} shows that it is very large,
classified as a ``giant'', with a radius $0.75\arcsec= 300$\,AU. 
A neighboring star lies only $1.1\arcsec$ (440AU) away, with a low probability
($<$2\%; \cite{kohler,reipurth}) of chance alignment.
Our SMA observations reveal substantial dusty disks around both
stars with masses of 0.066$\,M_\odot$ and 0.018$\,M_\odot$.
There is no evidence for any circumbinary material in the
HST or SMA data, which is not surprising given the large
separation.

Binary stars are important, not only because they make up a significant 
fraction of stars in the Orion Nebula \citep{petr,kohler,reipurth}
but also because at least 20\% 
of extra-solar planet discoveries are hosted by binary star systems
\citep{raghavan,desidera,eggenberger}.
However, millimeter wavelength detections of binary protostars
in Taurus-Auriga \citep{jensen03} and $\rho$\,Ophiuchus \citep{patience}
generally find disks only around the primary stars.
This agrees well with numerical models of core fragmentation which
predict a higher disk mass around the more massive star \citep{bate}.

\cite{rodriguez} found massive disks $\approx$ 0.01-0.05$\,M_\odot$
toward each of the binary protostars in L1551,
the range due to the uncertain contribution from free-free
emission at the long 7-mm wavelength of the observations.
These are deeply embedded Class 0 protostars, however, with substantial
circumstellar material and no clear dividing line between envelope and disks.
\cite{jensen03} detected two disks in the HK Tau
binary system but with low masses, $<\,0.0019, 0.003\,M_\odot$.
The binary proplyd 253-1536 stands out as the first
example of two optically visible stars each with sufficient
mass to form a Solar system\footnote{The triple system, UZ Tau EW studied by \cite{dutrey,jensen96} 
contains one star, UZ Tau E, with a disk mass high enough to potentially form a
solar system.  However, the stars of the accompanying 
binary UZ Tau W do not appear to have sufficently high individual disk masses.}
Their separation, $>$ 440 AU in projection, is large
enough that both the evolution of the disks and their prospects
for planet formation can be considered independently of each other
\citep{beckwith,desidera}.

The discovery of this binary disk system was serendipitous.
The disk around the optically brighter member of 253-1536
is too small to be seen in the glare of the star in the HST data.
Optical images reveal the presence of similarly sized disks in the
Trapezium cluster only through the presence of their photoevaporative tails.
If the 253-1536 system was closer to the cluster center,
both disks would be more apparent.
There are probably many other more small disks in Orion that
are not being photoevaporated and therefore not being detected by HST imaging.

The optically fainter star, 253-1536a has the larger and more massive disk.
It is spatially resolved in both the HST and SMA data. Unfortunately
it is impossible to meaningly constrain the surface density profile
without knowledge of the mid- and far-infrared spectral energy
distribution. \cite{smith} noted that the disk silhouette appeared
slightly de-centered from the stellar position. The effect would be
at the limit of the SMA resolution but is not seen.
The position angle of the disk in the HST image also appears
slightly smaller than the SMA image. These differences
are probably due to
lower density material that absorbs optical light but emits
little in the submillimeter. We speculate that an asymmetric
flaring of the disk, such as seen in 216-0939, could produce
this effect. 

As a binary system, the two circumstellar disks in 253-1536 formed at
the same time. Yet they have quite different masses and radii.
The stellar masses play a large role in the initial
disk conditions and subsequent evolution.
\cite{jensen03} noted that the HK Tau binary has a stellar mass
ratio close to unity and that other binaries with disk detections
only around the primary had higher mass ratios. 
Although the optically brighter star 253-1535b 
is known to be of spectral type M2.5 (Hillenbrand 1997),
the spectral type of 253-1535a with the larger disk, is unknown.
It is fainter by a factor of 40 in  H$\alpha$ \citep{reipurth},
by a factor of 6 at $2.2\,\mu$m \citep{kohler} and a factor
of 1.7 at $10.4\,\mu$m \citep{nielbock}.
The nearly edge-on disk obscures a considerable fraction of the
light at these wavelengths, however, and a spectrum is required
for definitive typing and to determine the stellar mass ratio in this system.
With this information, this system will be an important benchmark
for comparison with theories of disk formation and early evolution.

\section{Conclusions}

We have detected strong $880\,\mu$m emission from two silhouette
disks in M43, just north of the Orion Trapezium cluster.
The implied disk masses, 0.045$\,M_\odot$ and 0.066$\,M_\odot$,
are the largest yet discovered in this massive star forming region
and strengthen our conclusion from Paper I that Orion disks likely
had similar initial properties to those in the lower mass and less
crowded Taurus-Auriga and $\rho$\,Ophiuchus regions.

We have also found that each star in the binary system 253-1536
possesses its own disk. Only the larger disk is visible in HST
images but our SMA data, at comparable 0.2$\arcsec$ resolution,
reveal a 0.018$\,M_\odot$ disk around the optically brighter star.
Both disks in this binary have sufficient mass within 60 AU radius
to form planetary systems on the scale of our own but their
different masses and sizes demonstrate that the disks may
have formed with quite different initial conditions or that
their evolution is strongly dependent on additional parameters
than time alone.

Our SMA survey of circumstellar disks in Orion both in and beyond
the Trapezium cluster shows that about half the mass can be
lost in the outer edges of disks around stars within 0.2-pc
of $\theta^1$\,Ori\,C but large, massive disks can survive beyond 1-pc.
Further observations of a statistically representative sample of
disks at intermediate distances are necessary to determine the
sphere of influence of this O6 star more precisely. 
Additional observations are also required to better characterize
the upper end of the Orion disk mass distribution.
Finally, the high disk masses that we have observed would be
detectable out to 2-kpc and suggest that studies of other HII
regions could be fruitful.

\acknowledgments
This work is supported by the NSF through grant AST06-07710.

\begin{figure}[ht]
\vskip -0.9in
\plotfiddle{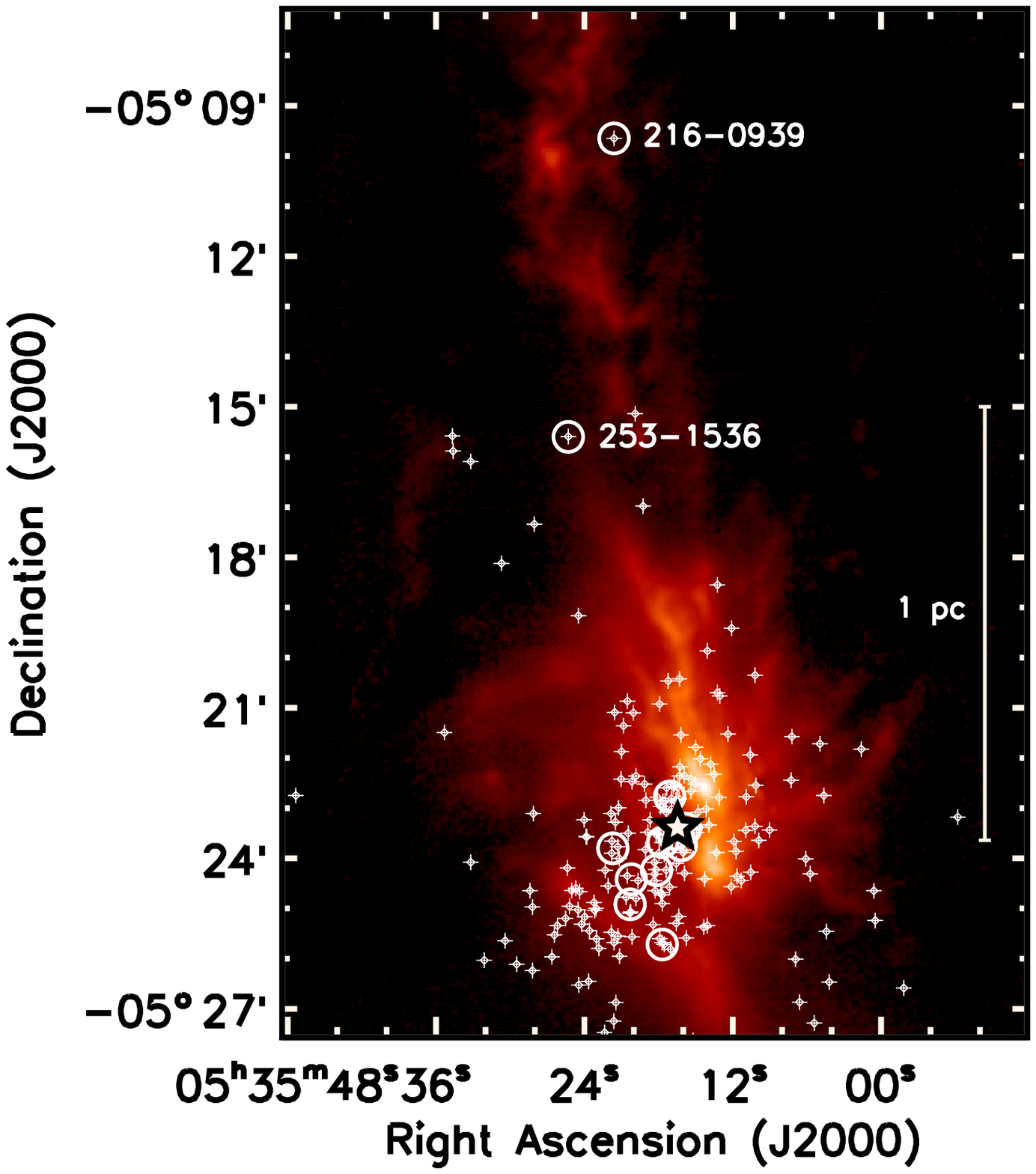}{-3.0in}{0}{625}{800}{-80}{20}
\vskip -2.75in
\caption{Large-scale view of the Orion Nebula at $450\,\mu$m.  Imaging was done
using SCUBA on the 15-m James Clerk Maxwell Telescope \citep{johnstone99}.
The crosses mark the location of HST-identified proplyds, and the
star shows the position of $\theta^1$\,Ori\,C, the most
massive star of the Trapezium cluster.
Solid circles represent the $32''$ primary beam of each SMA field.
The massive disks, 216-0939 and 253-1536 are clearly marked with their location.}
\end{figure}

\newpage
\begin{figure}[ht]
\vskip -1.0in
\epsscale{1.0}
\plotone{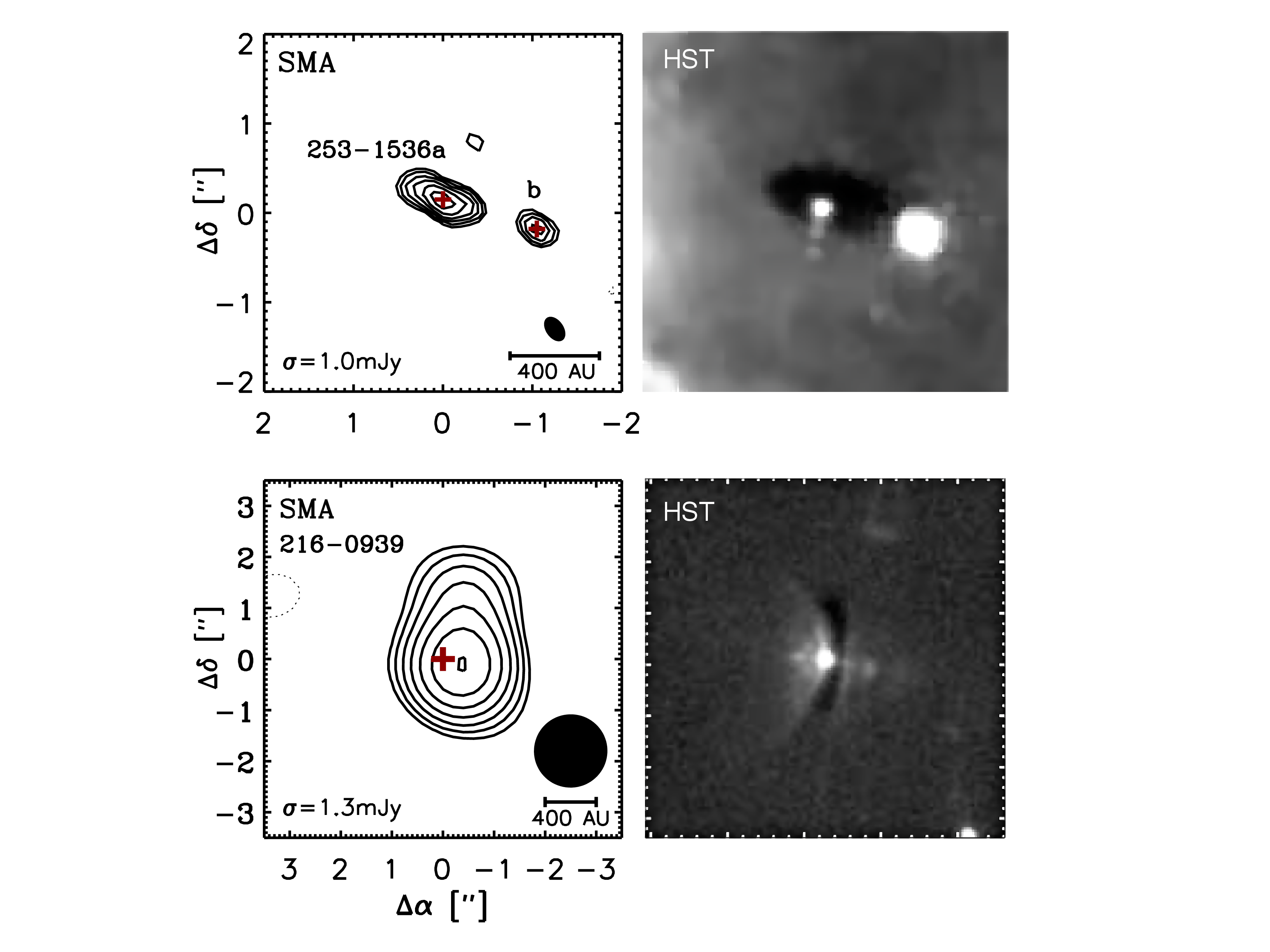}
\caption{Submillimeter Array images of the $880\,\mu$m continuum emission
from Orion disks 216-0939 and 253-1536a, 253-1536b.  HST H$\alpha$ discovery images 
were taken directly from \citet{smith} and are also shown using the same 
field of view: $7''\times 7''$ for 216-0939 and $4''\times 4''$ for 253-1536a and b.  
Contours begin at the 5$\sigma$ level, where $\sigma$ is the rms noise
level in the map, and is specified in the lower left corner.  Each step 
represents a factor of 1.5 in intensity. The synthesized beam size is shown 
in the lower right corner of each map.}
\end{figure}

\end{document}